# An analysis of the Internet of Things in wireless sensor network technologies

Harshit Poddar [1] [0000-0002-1307-6229] and Vansh Singh [2] [0000-0003-1328-6653]

[1] Vellore Institute of Technology, Vellore, Tamil Nadu, India
[2] Vellore Institute of Technology, Vellore, Tamil Nadu, India
[1] harshit10233@gmail.com
[2] Singhvansh1203@gmail.com

**Abstract:** Information may be accessed from a distance thanks to computer networks. Wireless or wired networks are also possible. Due to recent developments in wireless infrastructure, wireless sensor networks (WSNs) were developed. Activities or events occurring in the environment are monitored, recorded, and managed by WSN. Through a variety of routing techniques, data relaying is done in these systems. The fourth industrial revolution, or Industry 4.0, is defined as the integration of complex physical automation systems made up of machinery and devices connected by sensors and managed by software. This is done to boost the efficiency and reliability of operations. Industry 4.0 is viewed as a possibility because of industrial IoT, the concept of leveraging IoT technology in manufacturing. delivering, in an industrial setting, a means of connecting engines, power grids, and sensors to the cloud. In this essay, we'll try to comprehend how the Internet of Things (IoT) works in wireless sensor networks and how it might be used in various situations.

**Keywords:** IoT, Wireless sensors networks, industry 4.0, real-time data, security, adaptive.

## 1. Introduction

Each day lifestyles have modified considerably in all respects. the starting of wireless networking technology. The Internet of things (IoT) is in particular one of the quickest evolving technologies of the future. Multiple devices can be related in the physical world, which in truth adjusts each day life, inclusive of IoT. The want for communications in all locations and every time, particularly in fields with increased activity, is, therefore, increasing rapidly [1]

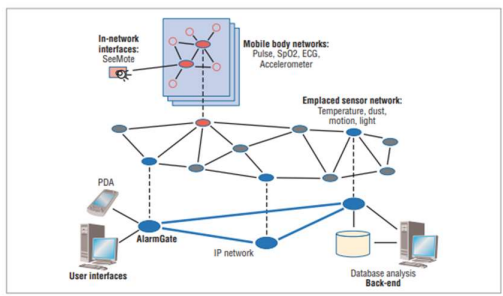

Fig 1. WSN with many levels of heterogeneity to support in-home or assisted living healthcare.

Many developments in the field of wireless sensor networks, telecommunication, and informatics have molded the realization of pervasive intelligence.

These functions require Higher data rates, enormous bandwidth, doubled capacity, minimal latency, and fast throughput are required for these operations. Given these new ideas, IoT has completely changed the world by enabling seamless connectivity between disparate networks (HetNets). IoT's ultimate goal is to bring plug-and-play technology that makes devices simple to use, accessible from a distance, and configurable for the user. The IoT technology is presented in this article from a high-level perspective, including its statistical and architectural trends, application cases, difficulties, and hopes for the future.

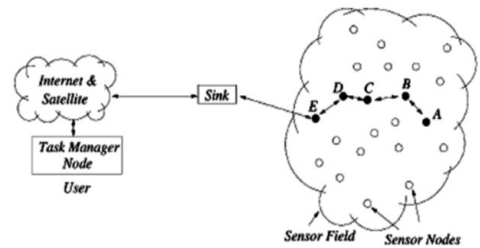

Fig 2. A standard wireless sensor network [5]

Real-time application has gained prominence among technocrats and researchers as a result of recent advances in the field of sensors. Researchers and engineers devised real-time wireless sensor network applications as a means of overcoming the obstacles presented by sensors (WSN). The end-user will receive input from the real-time sensors so that the information it has collected may be processed further. The performance of crucial applications that require constrained delay latency is specifically addressed by the real-time application. Real-time wireless communication is a WSN application area that is still in its infancy but has the potential to be a significant research area. apps that operate in real-time and can monitor, react quickly to user input, or influence an external environment.[2-3].

Wireless Sensor Networks (WSNs) have begun to catch researchers' attention because of the quick technical advancement of wireless technology and embedded electronics.

A typical WSN is made up of small devices called nodes. as well as nodes These nodes have an integrated CPU and limited memory. processing power, as well as some smart sensors Using these,

Sensors and nodes are used to monitor the environment. Humidity, pressure, heat, and vibration are all elements to consider.

A sensor interface is typically found on a node in any WSN. The computational unit, transceiver unit, and power unit are all interconnected. These are the units

That fulfills critical duties by enabling nodes to interact among themselves to transfer data acquired by their sensors

WSN can be deployed on land, underwater, or underground, depending on the scenario. There are several varieties of WSN, notably terrestrial WSN, subsurface WSN, underwater WSN, multimedia WSN, and mobile WSN. [5,7,8].

## 2. How do wireless sensor networks function?

Sensor nodes are the small, inexpensive components that make up wireless sensor networks. It might be little or massive. Because of this, build the wireless sensor network using sensor nodes.

As a result, the sensor nodes are crucial to the complete functioning of the sensor network. These nodes vary in size and rely on this since different sensor node sizes perform efficiently in different sectors. A radio transceiver for producing radio waves, a variety of wireless communication devices, as well as a ready energy source like a battery, are all features of the sensor nodes used in wireless sensor networking.

The whole network operated simultaneously utilizing several types of sensors and utilized the multi-routing algorithm, sometimes known as the wireless AD HOC network.

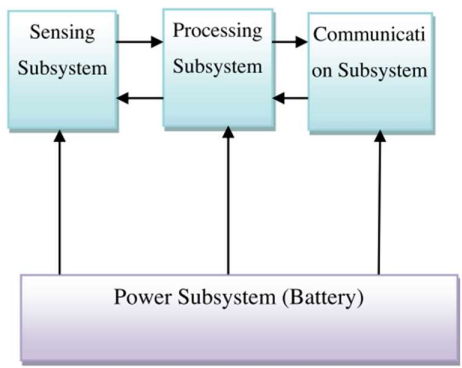

Fig 3. common sensor node architecture for the Internet of Things.

It is depicted in Fig. 3. Secondary components for the aforementioned parts include filters, amplifiers, transducers, comparators, etc. Data from the workplace is collected or sensed by the sensing device. The communication unit sends data at the BS (base stations), while the processing unit performs various data manipulation activities, such as data gathering, and the power unit, which is often a battery-limited one, supplies power to all other devices.

## 3. IoT implementation in WSN

Now in the above section, we have seen what is the meaning of WSN and how does wireless sensor networks (WSN) work?

Before that first, let us understand different types of sensors and their challenges. Bellow tables. Give an enormous idea of the different types of sensors available and the challenges of using those sensors.

Table 1. Comparison of selective types of sensors based on their characteristics [17]

| Type of sensors | Sensitivity | Non-linearity error | Hysteresis error | Resolution | Accuracy | Offset and Zero drift |
|---|---|---|---|---|---|---|
| Resistive Sensor | Resistive material and electrical properties | Straight line and an electrical output curve | Two curves of electrical output | Resistivity of the material | Actual value and the output value of electrical property | Electrical value and response curve intersect point |
| Inductive and magnetic sensor | Electrical inductance and magnetic field and displacement property | Straight line and displacement or velocity output curve | Two curves of displacement or velocity output | Electrical inductance and magnetic field in the material | Actual value and the output value of displacement or velocity property | Displacement value and response curve intersect point |
| Piezoelectric sensor | Resistive material and conductive particles | Straight-line to the resistance output curve | Two curves of resistance output | Force or pressure on the resistivity of the material | Actual value and the output value of resistance property | Resistance value and response curve intersect point |
| Acoustic sensor | Light intensity and velocity or amplitude of the wave | Straight-line to the velocity or amplitude of the wave output curve | Two curves of velocity or amplitude of wave output | Sound intensity | Actual value and the output value of velocity or amplitude wave property | Velocity or amplitude of wave value and response curve intersect point |
| Potentiometer sensor | Resistive material and voltage change in our circuit | Straight-line to the voltage output curve | Two curves of voltage output | Resistivity of the material | Actual value and the output value of voltage property | Voltage value and response curve intersect point |
| Thermoresistive sensors | Resistive material and temperature | Straight-line to the temperature output curve | Two curves of the temperature output | Resistivity of the material | Actual value and the output value of temperature property | Temperature value and response curve intersect point |
| Optoresistive sensors | Photo resistive material and light intensity | Straight-line to the light intensity output curve | Two curves of light intensity output | Photoresistive material | Actual value and the output value of light intensity property | Light intensity value and response curve intersect point |
| Capacitive sensors | Electrical charge of the material and voltage change across two plates | Straight-line to the voltage output curve | Two curves of voltage output | Electrical charge of the material | Actual value and the output value of voltage property | Voltage value and response curve intersect point |

Table 2. Analysis of sensors based on methodology, application, and problems [17]

| Types of sensors | Methodology | Applications | Research issues and challenges |
|---|---|---|---|
| Resistive Sensor | It detects physical properties by relating it to change resistance. | Force resistive sensor, emitting diode display controller | Dynamic light scattering, nanoparticle tracking analysis, tunable resistive pulse sensing and nanoparticle tracking analysis. |
| Inductive and magnetic sensor | It relates the physical properties to electrical inductance and the changing magnetic fields. | Search coil magnetometer, inductive proximity sensor. | To obtain electrical energy from the kinetic energy, resonant coupling and near field wireless power transfer. |
| Piezoelectric sensor | It detects changes in resistivity when squeeze, press or deformed that lead to changes in the electrical properties. | Car alarms, traffic camera, touching display, digital blood pressure monitoring, atmospheric pressure. | Continuous contact-free monitoring of respiration and heart, utilization of supply sustainable. |
| Acoustic sensor | It detects changes in the properties of the material affects the velocity or amplitude of wave signals. | Widely used in the medical, technological, scientific application. | Distribution of task and multi-robot coordination, underwater coverage problem for sensor networks. |
| Potentiometer sensor | It detects changes in the position of wiper gives change in voltage coming through pins. | Electromotive force, variable resistor, adjusting electronic circuits. | Widespread use in sport and biomechanics studies. |
| Thermoresistive sensors | It detects material changes its resistance in response to changes in temperature. | Thermistor, diodes, digital temperature sensor. | Development of thermal flow sensors, Multiparameter detection in microflows and nanoflows. |
| Optoresistive sensors | It detects changes in photoresistive material that respond to change in light intensity. | Light controlled variable resistor, clock radios, alarm devices, solar streetlamps, solar road studs. | To design optically powered low power and low-cost electronic sensors. |
| Capacitive sensors | It defines the amount of charged stored in the capacitor is equal to the capacitance times the voltage. | Level control of liquids, level control of solids and pile up control. | Presence detection of floating targets, detection of the dielectric floating rotor in an angular encoder, detection of displacements over small range. |

Now in this section let us see how it is used in WSN and what the different types of WSN are there. Wireless sensor organization (WSN) is vital in PC organizing for looking through the region and in data collection. WSN tracks down its application in a few regions, information capacity, and observing. "Sound Surveillance System" was the name of the first fully developed WSN (SOSUS). It was utilized to recognize the danger posed by submerged submarines." [14].

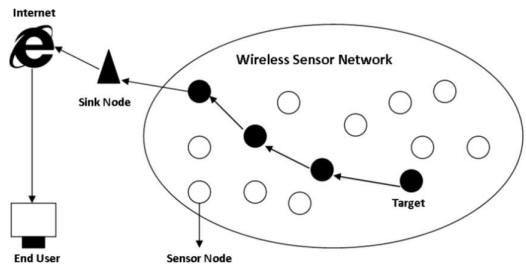

Fig 4. architecture of WSN system [14].

WSN comprises a client, an interconnected spine, and a sensor hub. The figure presents WSN Arrange engineering comprising sensor hubs, which are utilized in environmental monitoring. Within the setting of remote communication, the sensor hubs communicate with each other and send the handled information to the sink hub. All the hubs send information to the sink hub, which is further sent to clients through the web.

Now let us see what the different types of wireless sensor networks are there. The below figure. depicts the different types of WSNs available for different purposes.

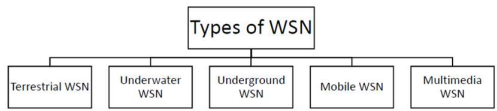

Fig 5. Different types of WSNs [14].

1) Mobile WSNs: - Power moves and automatically links to the environment, which is a benefit of a mobile WSN. Mobile sensors interact and link to computers. Mobile sensors may gather data from a large region or data from other nodes or sensors. The current state, coverage area, navigation, repositioning, and maintenance of mobile WSNs are their primary drawbacks [14].

2) Underground WSNs: - A sensor node buried beneath the surface of the earth to gather data on the conditions underground is known as an underground WSN. The limited battery power of WSN is a disadvantage since it is exceedingly challenging to recharge or replace.

3) Multimedia WSNs: - To enable tracking and monitoring of multimedia events, such as audio, video, and image, multimedia wireless sensor networks are used. Low-cost sensor nodes with cameras and microphones are present in these networks. For data retrieval, data compression, and data correlation, these multimedia WSN sensory nodes are coupled through a wireless network.

4) Terrestrial WSNs: - Terrestrial WSNs involve different small sensor centers. These center points are heedlessly conveyed in a specific zone from where a promotion hoc orchestrate is used for correspondence between the centers. The structured model takes into account grid placement, 2D and 3D optimal placement, and placement models [14].

5) Underwater WSNs: - As we know the earth is covered with approximately 75% of water. So, in this several sensor nodes and submerged cars makeup networks. These sensor nodes' data are collected by autonomous underwater vehicles. Long propagation delays, low bandwidth, and sensor failures are difficulties in underwater communication.

As we have seen different types of WSNs, now let's see their applications. As there are many different types of applications are there here, we will try to take some application and explain it in brief.

### 3.1 Smart Grid

So, before starting the application of WSNs in the smart grid first, let us understand what is the meaning of the smart grid?

Using two-way digital communication, a smart grid is an electrical distribution system that is based on digital technology. To increase productivity, cut costs and consumption of energy, and increase the transparency and dependability of the energy supply chain, this system enables monitoring, analysis, control, and communication across the supply chain and we called it a smart grid.

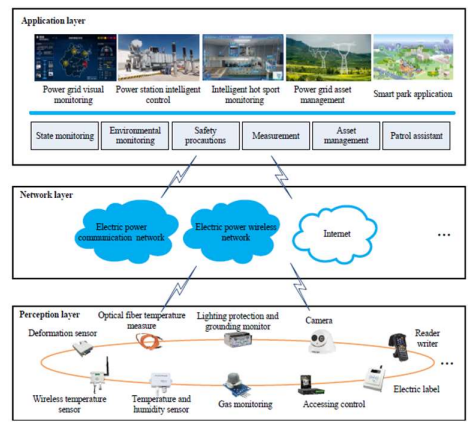

Fig 6. The general design of the smart grid wireless sensor network [15]

Above figure. tell us about the overall construction of the wireless sensor networks in the smart grid.

The expected characteristics of WSN used in the smart grid are listed below [16]:

1) The ability to communicate: - The entire connection distance of sensor nodes in a smart grid is constrained. Unpredictable weather conditions, including wind speed, rainfall, etc., can seriously damage network performance.

2) Multi-hop communication: - The WSN network's lifespan is extended and power consumption is decreased by using multi-hop communication. A node may occasionally need to connect to a non-direct neighbour node utilizing gateways and routers to send data with the assistance of intermediary nodes.

3) Energy for the battery: - Nodes will get deactivated if the power supply were lost. The protocols and algorithms employed should also decrease the system's energy usage.

4) Computational power: - The smart grid's sensor computing power is constrained, and memory and storage space is also at a minimum. The cost of the sensor node is more than that of changing the battery or disassembling the unit.

5) One's organization: - WSNs have self-organizing capabilities, they don't require a set-up network architecture. Sensor nodes need to cooperate effectively and perform well.

### 3.2 Environmental Surveillance

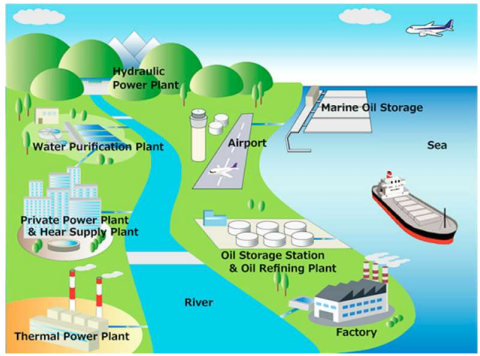

Fig 7. Environmental monitoring system based on WSN [14]

WSNs can be used for real-time data monitoring or environmental surveillance. Nowadays environmental care has become an important priority for the whole world because of the increase in global warming different climatic changes are happening and because of these changes, we can use different types of WSNs to get and surveillance real-time environmental data. So, to set up this station we will require a weather station with a large battery capacity for long-lasting working types of equipment, the networks should be performed simply and should perform the expected operations and the radio transmission signal should be good. Like, this we can try to set up the environmental surveillance

Table 3. Examples of prototypes for environmental monitoring [14].

| Application | Sensors | City Deployed in |
| --- | --- | --- |
| Waste Removal | Global System for Mobile (GSM) Tracking Sensors | New York, Seattle |
| City See $CO_2$ Monitoring | $CO_2$, Temperature, Light Sensors | Wuxi, China |
| Air Pollution Maps | $NO_2$, CO, $O_3$, Ultrafine Particles (UFPs), Temperature, Humdity | Zurich, Switzerland |
| Air Quality | $NO_2$ Sensors | Amsterdam, The Netherlands |

### 3.3 Health Care Industries

WSNs are widely used in the health care industry. Remote health monitoring is one of the open and well-liked uses of the health services offered by the Internet of Things (Telehealth). Patients no longer always need to go to the hospital or even the emergency department thanks to the deployment of IoT. In addition to lowering expenses and reducing the necessity for hospital visits, this service also enhances patients' quality of life by relieving them of the inconvenience of hospital travel. Patients who have trouble walking or moving about might benefit greatly from even a small convenience, especially if they use public transportation [18].

Detecting and Treating Chronic Diseases: As life expectancy has grown, so has the prevalence of chronic health issues. Fortunately, significant advancements have been made in the treatment of many issues, and IoT adoption can significantly affect many of them. Different methods of recognizing and treating chronic illnesses can be introduced directly through wearable technology, next-generation analytical tools, and mobile devices.

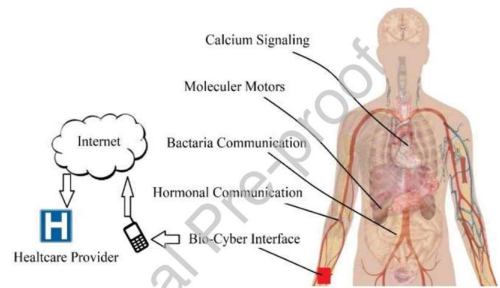

Fig 8. As an example, consider the Bio-IoT network design. [18]

Personalized medicine, identifying and treating chronic illnesses, supported living, and care for the elderly are just a few of the areas that may use better when it comes to IoT in WSN applications in the health care system. IoT has several uses in WSNs, including home automation, the detection of chemical and biological attacks, patient monitoring in the medical field, the detection of landslides, etc.

## 4. Difficulties faced in WSN

Different heterogeneous relics introduced and imparted in various settings achieve IoT's intricacy and make the sending of safety instruments considerably more convoluted. Existing WSN security research offers answers for abstract issues without considering the effect of the IoT standards and elements as analysed in this archive.

### 4.1. Energy

Sensor elements are a key element for observation or tracking systems. Sensor nodes are inexpensive, lightweight, and battery and power limited. The power supply is a critical factor in the lifespan of the sensor. Energy is consumed for node operations cherish detection, data collection, and network operations such as data communications through various communication protocols. Batteries are small and need to be replaced or recharged, this is not always possible. There are also many strategies for power generation, but they cannot eliminate the need for power management. Life's biggest challenge is limited battery management by planning and implementing various low-power hardware and code protocols for WSN [27]

### 4.2. Real-time management

It is a challenging problem for resource-controlled sensor networks. In such instances, a smart data-driven middleware design is required for the IoT system to send factual info only when essential, as well as an effective service gateway design to minimize the quantity of data to be transferred by constantly reviewing user data. greater than threshold reading [27-28]

### 4.3. INTEROPERABILITY

One of the tasks that the WMSN protocols may offer is compatibility and interoperability amongst heterogeneous networks, including multiple wireless/wired technologies. It is required to give the capacity to interoperate several networks without impacting the complexity of software and hardware components for this purpose. It completely employs the same standard and compatibility with various wired/wireless technologies via commercial gateways.

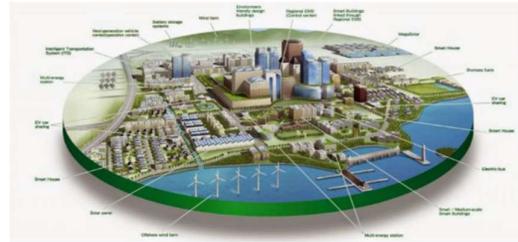

Fig 9. WSN with IoT practical application [26]

### 4.4. SECURITY AND PRIVACY

Wireless networks can be readily intercepted, modified or replaced, destroyed, introduced, or damaged by external/internal attackers due to the weakness, susceptibility, or non-existence of security systems. For example, a malevolent, dishonest false user may continue to submit frequent requests to the network, blocking legitimate users who have access to touch data from accessing it. As a result, one of the most important components of WMSN is security.[29] Depending on the program's complexity, WSNs can provide data confidentiality, verification, fairness, and usability without Internet access. Physical activity near the WSN is required by the attacker to add malicious nodes to the present network or to stop or catch them. However, the introduction of WSNs on the internet allows attackers from all over the world to carry out their harmful operations. As a result, the WSNs must firmly handle the difficulties that arise from this Internet connection, such as malware and others. Current WSNs supply a key and a specific effective gateway to provide efficient security. However, because of restricted computer power, energy, and memory, it is hard to reproduce the same security system. In comparison to other types of Internet networks, sensor nodes for Greater secrecy have not yet adopted RSA-1024-key length encryption. Better security systems must also include the resource limitations that now exist to prevent different Internet-related assaults. [28].

### 4.5. Localization

Using optimization techniques such as PSO, firefly optimization, etc., the localization issue may be successfully resolved. Some of the problems that this issue raises must be addressed, though. For instance, in a sensor network, beacon nodes or another effective strategy must anticipate and detect threats. Hybrid algorithms' time complexity increases when we apply them to problems, hence the algorithm's time complexity must also be decreased.[30].

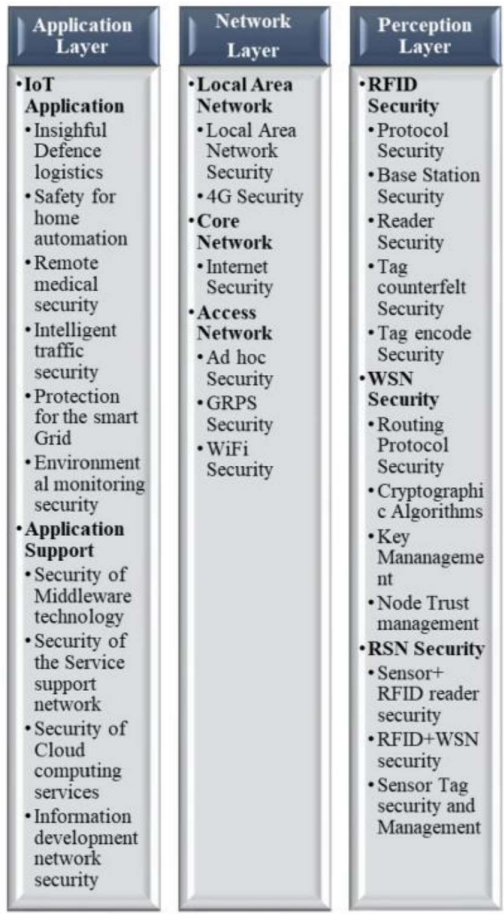

Fig 10. IoT security architecture in particular [26]

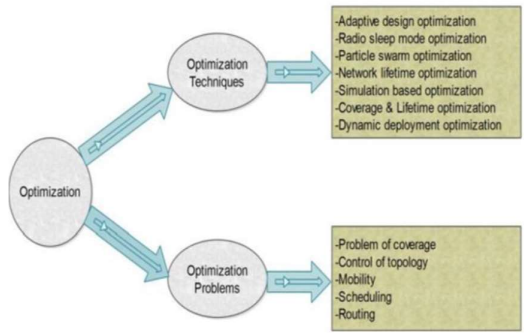

Fig 11. WSN optimization approaches and problems [31]

### 4.6. Low-Cost Sensor Devices, Hardware, and Resource Constraints

The IoT objects typically suffer from a serious lack of resources (e.g., power, storage, etc.). Their storage and computing capacities are unable to handle sophisticated processes. Additionally, they frequently use depleted batteries and are typically battery-powered. Power is regarded as the most important restriction among the others since sensing and processing must be completed in a short amount of time. Complex processes are not possible because of computational and storage limitations as well as battery power limitations, among other factors. Resource management systems distribute the jobs across the nodes through process migration.

The nodes in sensor networks are often diverse for greater network application; they have various energies, functionalities, communication capabilities, and so forth.[33]

### 4.7. Challenges Based on Hardware

As the controller is stated to be a centralized device that serves as the network's brain, it is thought that controllers are the main source for attackers. There are many traditional methods, and the newer assaults behave in a way that makes it easy to deactivate the controller. For instance, a Classic DDoS attack may target the controller. Aggressors can produce a huge volume of fraudulent packets and send them all at once to the switches. Switches that would afterward provide the least number of requests for fraud transmission to the controller or an equal amount of requests would regard all fraud packets to be brand-new. Thus, the controller's processing resources will be used in a significantly shorter amount of time in this manner. Additionally, adversaries have thought of the transition as being weak. According to the OpenFlow protocol, switches would likely be changed into the independent mode or the fail-secure mode because switches have the least amount of execution in hardware resources and can therefore be initially attacked through the channel of communication between controllers and switches. The network's performance would unquestionably suffer as a result.[34]

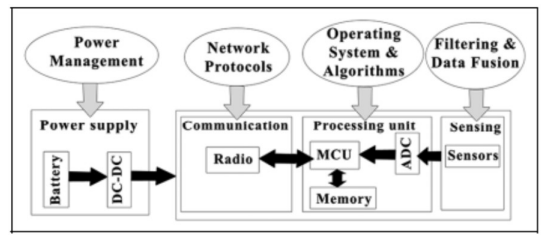

Fig 12. Sensor node block diagram [35]

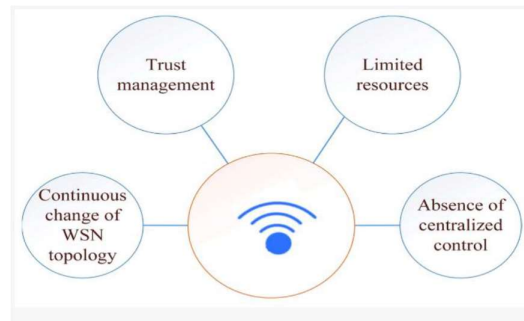

Fig 13. WSN's Main security challenge [36]

### 4.8. Availability

Compromise nodes are necessary for access to WSNs. Incorporating an encryption method for WSN security may cost extra. The utilization of auxiliary communications and code update and reuse are only two of the effective strategies researchers have created to achieve their objectives. In addition, methods for accessing the data have been developed. Therefore, availability is crucial to preserving the operational services of WSNs. It also helps to maintain the network as a whole until its completion.[27]

### 4.9. Future trends

The use of Edge and Fog computing in conjunction with optimization techniques to reduce latency and assure security is a potential future trend. Since these algorithms are the package of addressing issues in multiple aspects, such as how the ML method in WSN/IoT handles prediction-based, error-based problems and optimization to solve local minima/maxima-based problems, the study also needs to incorporate more hybrid algorithms to answer a difficulty. Any wireless network has issues similar to those mentioned in this article. In light of the same problems and difficulties, the research may be broadened to include extended versions of IoT networks and fog computing in IoT.

## 5. Conclusion

The wireless sensor network is crucial to wireless networks because it can track changes in the physical and environmental conditions that older networks are unable to. In this article, we explore the problems, characteristics, and protocol stack of the wireless sensor network, as well as how it differs from the standard network. Advances in computer technology have led to the expansion of WSNs, which have existed since the required characteristics. In recent years, IoT-based WSN systems have attracted a lot of interest. Nonetheless, these systems suffer from limited bandwidth, power, and resources during point-to-point transmission. The existing works defining the role of IoT in WSN challenges and applications are reviewed in this paper.